\pdfoutput=1

\documentclass[10pt,conference]{IEEEtran} 
\usepackage{cite}
\usepackage{mathtools}
\usepackage{amsmath,amssymb,amsfonts}
\usepackage{algorithmic}
\usepackage{graphicx}
\usepackage{textcomp}
\usepackage{xcolor}
\usepackage{physics}
\usepackage{enumitem}
\setlist[itemize]{noitemsep, topsep=0pt}
\setlist[itemize]{leftmargin=*}
\usepackage{caption} 
\usepackage{hyperref}
\newcommand{\subparagraph}{}
\usepackage[compact]{titlesec}
\titlespacing{\section}{0pt}{5pt}{1pt}
\titlespacing{\subsection}{0pt}{*0}{*0}
\def\BibTeX{{\rm B\kern-.05em{\sc i\kern-.025em b}\kern-.08em
    T\kern-.1667em\lower.7ex\hbox{E}\kern-.125emX}}

\usepackage{etoolbox}
\author{\IEEEauthorblockN{Jasmine Sekhon}
\IEEEauthorblockA{
\textit{University of Virginia}\\
Charlottesville, VA USA \\
\href{js3cn@virginia.edu}{js3cn@virginia.edu}}
\and
\IEEEauthorblockN{Cody Fleming}
\IEEEauthorblockA{
\textit{University of Virginia}\\
Charlottesville, VA USA \\
\href{cf5eg@virginia.edu}{cf5eg@virginia.edu}}
}
\title{Towards Improved Testing For Deep Learning}

\makeatletter
\patchcmd{\@maketitle}
  {\addvspace{0.5\baselineskip}\egroup}
  {\addvspace{-1.5\baselineskip}\egroup}
  {}
  {}
\makeatother

\begin{document}
\maketitle
\begin{abstract}
The growing use of deep neural networks in safety-critical applications makes it necessary to carry out adequate testing to detect and correct any incorrect behavior for corner case inputs before they can be actually used. Deep neural networks lack an explicit control-flow structure, making it impossible to apply to them traditional software testing criteria such as code coverage. In this paper, we examine existing testing methods for deep neural networks, the opportunities for improvement and the need for a fast, scalable, generalizable end-to-end testing method. We also propose a coverage criterion for deep neural networks that tries to capture all possible parts of the deep neural network's logic. 
\end{abstract}

\begin{IEEEkeywords}
deep neural networks, whitebox testing, coverage criterion 
\end{IEEEkeywords}

\section{Introduction}\label{section:1}
Deep Neural Networks, or DNNs, are increasingly being used in diverse applications owing to their ability to match or exceed human level performance. The availability of large datasets, fast computing methods and their ability to achieve good performance has paved way for DNNs into safety-critical avenues such as autonomous car driving, medical diagnosis, security, etc. The safety-critical nature of such applications makes it imperative to adequately test these DNNs before deployment. However, unlike traditional software, DNNs do not have a clear control-flow structure. They learn their decision policy through training on a large dataset, adjusting parameters gradually using several methods to achieve desired accuracy. Consequently, traditional software testing methods like functional coverage, branch coverage, etc. cannot be applied to DNNs, thereby challenging their use for safety-critical applications.

A lot of recent work, discussed in \ref{section:3}, has looked into developing testing frameworks for DNNs. These methods suffer from certain limitations, as discussed in \ref{section:4}. In our work, we intend to make an effort to overcome these limitations and build a fast, scalable, efficient, generalizable testing method for deep neural networks. In \ref{section:5}, we propose a coverage criterion for feed forward deep neural networks that tries to capture the DNN logic to a greater extent by incorporating inter-layer and intra-layer relationships. 

\section{Background}\label{section:2}
Deep neural networks are neural networks with multiple hidden layers between the input and output layers. Unlike traditional software programs, where the program logic has to be manually described by the programmer, deep neural networks are capable of learning rules by training on a large dataset. Today, DNNs are used in easy to complex tasks, such as image  classification\cite{Y.LeCunLenet-5Networks}, medical diagnosis and end-to-end driving in autonomous cars\cite{Bojarski2016EndCars}.
The safety-critical nature of such applications makes it important to assure correctness, to avoid fatally incorrect behavior and obtain performance benefits from DNNs safely. 

\begin{figure}
    \centering
    \includegraphics[width = 0.5\textwidth]{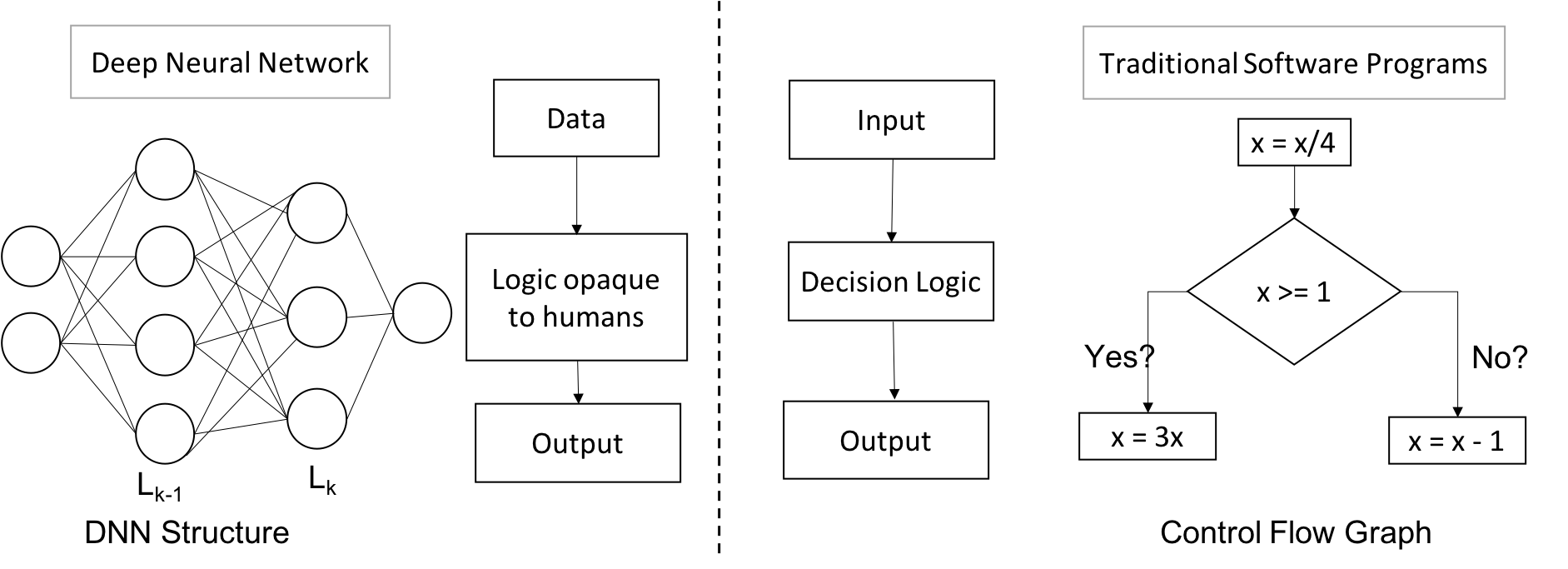}
    \caption{The internal logic of a deep neural network is opaque to humans, as opposed to the well laid out decision logic of traditional software programs.} 
    \label{fig:1}
    \vspace{-0.5em}
\end{figure}

\begin{figure}
    \centering
    \includegraphics[width = 0.4\textwidth, height = 0.2\textwidth]{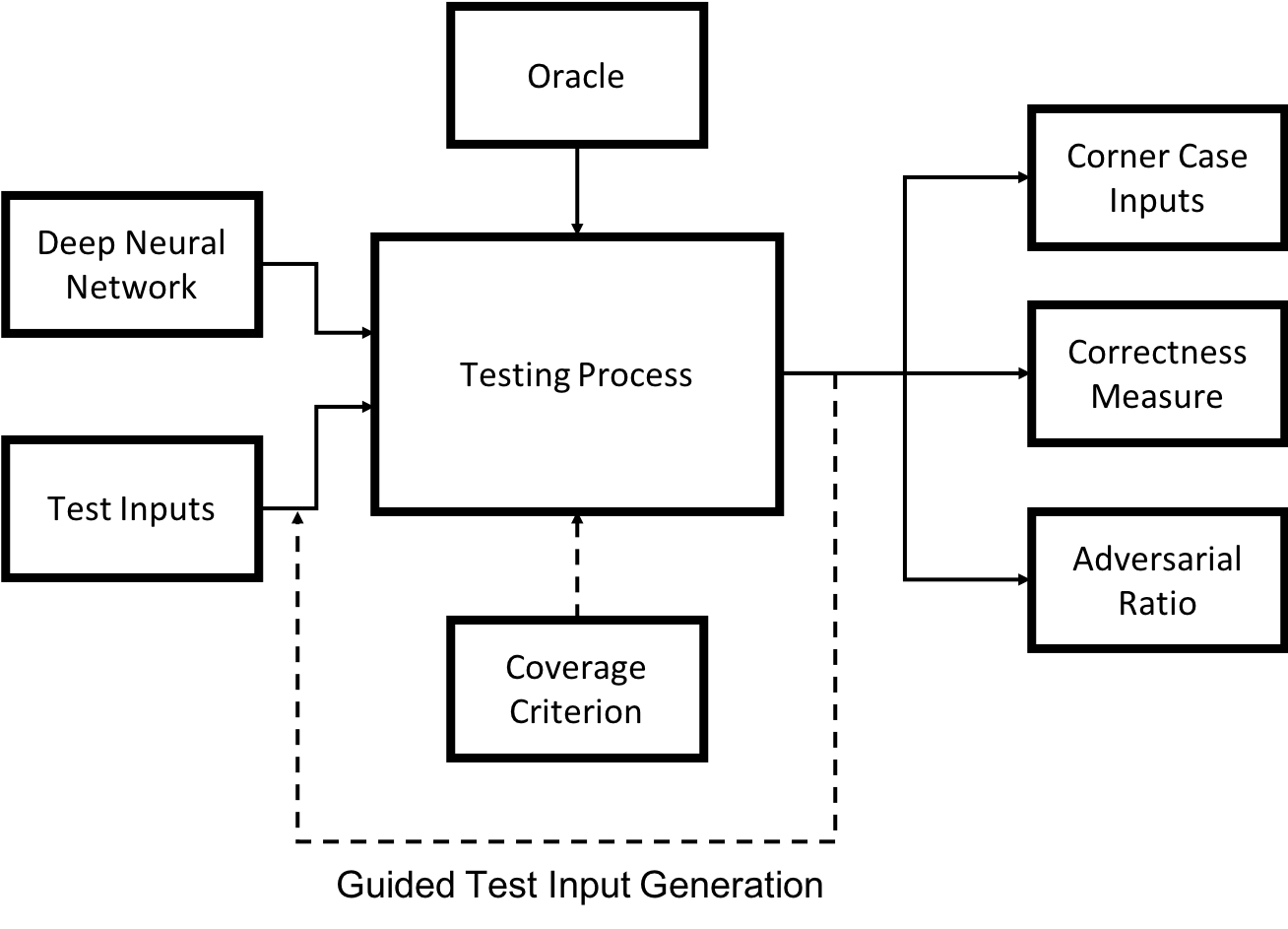}
    \caption{A high-level representation of most existing DNN testing methods. }
    \label{fig:2}
    \vspace{-0.5em}
\end{figure}

Traditional software testing methods fail when applied to DNNs because the code for deep neural networks holds no information about the internal decision-making logic of a DNN, as shown in Figure \ref{fig:1}. DNNs learn their rules from training data and lack the control-flow structure present in traditional software programs. Therefore, traditional coverage criterion like code coverage, branch coverage, functional coverage, etc. cannot be applied to deep neural networks. 
A high-level representation of most existing whitebox testing methods for DNNs is shown in Figure \ref{fig:2}. The inputs to the testing process are the DNN, the test inputs, and a coverage metric to ensure that all parts of the program logic have been tested. An oracle decides whether the behavior of the DNN is correct for the tested inputs. Further, a guided test input generation method may be used to generate test inputs that have \emph{greater coverage} and which \emph{uncover greater corner case behavior}. The output of existing testing methods is usually either a measure of system correctness or adversarial ratio. 
\section{Prior Literature}\label{section:3}
Testing methods for deep neural networks have normally followed a black-box approach, until recently, when DeepXplore\cite{Pei2017DeepXplore:Systems} proposed the first white-box testing method for DNNs. The method proposed by \cite{Narodytska2016SimpleNetworks} involves randomly searching around a given input for changes that cause misclassification. Many other approaches involve generating adversarial examples by perturbing an input slightly to induce incorrect behavior, which is checked for manually. However, these black-box approaches are completely unguided in terms of the absence of a coverage criterion and overlook the internal logic of a DNN. In DeepXplore\cite{Pei2017DeepXplore:Systems}, the authors introduced the concept of neuron coverage as a coverage metric for testing DNNs. They also proposed using multiple implementations for the same task as an oracle to avoid manual labeling effort. Further, DeepCover\cite{Sun2018TestingNetworks} proposes several criteria for testing DNNs, inspired by modified code/decision coverage for software testing. Their coverage criteria take into account the condition-decision dependence between neurons of consecutive layers. Another recent approach, DeepMutation\cite{Ma2018DeepMutation:Systems} is the first source-level mutation testing technique that proposes a set of model-level mutation testing operators that directly mutate on deep learning models without a training process. DeepCT\cite{Ma2018CombinatorialSystems} uses a combinatorial-testing inspired coverage criterion which guides an exhaustive search for test inputs that activate neurons in a layer-wise manner. 
\section{Opportunities for Improvement}\label{section:4}
\subsection{\textbf{Why do we need a better coverage criteria?}} 
Coverage criteria for traditional software programs, such as code coverage and branch coverage check that all parts of the logic in the program have been tested by at least one test input and all conditions have been tested to independently affect the entailing decisions. On similar lines, any coverage criterion for deep neural networks must be able to guarantee completeness, that is, it must be able to ensure that all parts of the internal decision-making structure of the DNN have been exercised by at least one test input. 

A typical feed-forward deep neural network contains multiple nonlinear processing layers with each hidden layer using the output of the previous hidden layer as its input. Each layer consists of multiple \textit{neurons}. A \textit{neuron} is a computing unit, loosely patterned on the neurons in the human brain, which fires/activates when it receives sufficient stimuli or input. 
Mathematically, if $L_{k-1}$ and $L_{k}$ denote two consecutive layers of this DNN (Figure \ref{fig:1}):
\begin{equation} \label{equation1}
n_{i,k} = \phi_{k}\Big(\delta_{i,k} + \sum_{1{\leq}j{\leq}N_{k-1}}(w_{j,i} {\cdot} n_{j,k-1})\Big)
\end{equation}
where:
\begin{itemize}
    \item $n_{i,k}$ denotes the value of the $i$th neuron of $k$th layer,
    \item $\phi_{k}$ denotes the activation function of the $k$th layer,
    \item $\delta_{i,k}$ denotes the bias for node $n_{i,k}$,
    \item $N_{k-1}$ denotes the number of nodes/neurons of layer $L_{k-1}$, and
    \item $w_{j,i}$ denotes the weight of the connection between the $j$th neuron of layer $L_{k-1}$ and $i$th neuron of layer $L_{k}$
\end{itemize}
Therefore, along with the value of each neuron having an independent effect on the activation of neurons in the next hidden layer, the combinations of values of neurons in the same layer also affect the value of neurons in the next layer. Any coverage criterion for deep neural networks must be able to capture both of these factors. Further, the coverage criterion should be scalable to larger-sized real-world DNNs and different network architectures. The coverage criteria proposed by previous works suffer from several limitations: 
\begin{itemize}
    \item Neuron coverage\cite{Pei2017DeepXplore:Systems} measures the parts of the DNN's logic exercised by the test inputs based on the number of neurons activated by the input. However, it is not able to thoroughly account for all the possible behaviors that a DNN could exhibit. Experiments by \cite{Sun2018TestingNetworks} were able to prove that neuron coverage is fairly easy to achieve and 25 random test inputs are able to achieve close to 100\% neuron coverage. Further, we observed that corner case behavior can be found beyond 100\% neuron coverage. Our experiments\footnote{All results for neuron coverage were obtained by running the DeepXplore code \url{https://github.com/peikexin9/deepxplore/tree/master/MNIST} for image manipulation=light and best-performing parameters: $\lambda$\textsubscript{1}=1, $\lambda$\textsubscript{2}=0.1, steps=10, grad\_iterations=1000, threshold=0} found that in certain cases of model architecture, for instance LeNet-1 used for MNIST handwritten digit classification, 100\% neuron coverage can be obtained with two test inputs, because for most test inputs, the neurons are always fired/activated. Therefore, neuron coverage is a fairly coarse and insufficient criterion for coverage in DNNs.
    \item DeepCover's\cite{Sun2018TestingNetworks} coverage criteria take into consideration the condition-decision dependence in adjacent layers of a DNN. Apart from their method being tested on relatively small networks, it assumes the DNN to be a feedforward, fully-connected network and cannot generalize to architectures like RNNs, LSTMs, attention networks, etc. Such methods do not consider the context of a neuron in its own layer, and the combinations of neuron outputs in the same layer.
    \item DeepCT's\cite{Ma2018CombinatorialSystems} combinatorial testing inspired coverage criterion determines the fraction of logic exercised by a test input in terms of the fraction of neurons activated in each layer. It does not consider the inter-layer relationships within a DNN, and has not been verified to scale to real-world DNNs with different kinds of layers. 
\end{itemize}
\subsection{\textbf{Why do we need better test input generation?}} 
Generating or selecting test inputs in a guided manner usually has two major goals - maximizing the number of uncovered faults, and maximizing the coverage. \cite{Pei2017DeepXplore:Systems} introduces a joint optimization based test input generation method, in which an existing test input is modified (using image manipulations) recursively until a test input causing differential behavior is found. \cite{Tian2017DeepTest:Cars} uses a similar greedy search technique in which random transformations are applied until an appropriate test input is found. Such test input generation methods suffer from some major drawbacks: 
\begin{itemize}
    \item The iterative process of manipulating an existing test input until a test input that satisfies the criterion is found, has considerable time per execution.
    \item The number of test inputs that actually cause an increase in coverage and/or an increase in the number of uncovered corner case behaviors are fairly low in comparison to the sum of total number of tested and generated inputs. 
\end{itemize} 
\subsection{\textbf{Why do we need a better oracle?}}
Testing for the correctness of a DNN requires the presence of ground truth (oracle), that decides if the behavior is correct. The existing oracles for testing DNNs suffer from several limitations:
\begin{itemize}
    \item The most straightforward way in data-driven schemes like DNNs is by collecting as much real-world data as possible and manually labeling it to check for correctness. However, such a process requires a lot of manual effort. 
    \item In multiple DNN implementations\cite{Pei2017DeepXplore:Systems} as an oracle, multiple implementations for the same task are compared, and differential behavior is labeled as a corner case behavior. However, we observed that this method erroneously classifies certain corner case inputs as correct behaviors because the labels predicted by all the implementations are similar and misclassifies several correct inputs as corner-case behaviors. Also, this method is only valid in applications that have several existing high-accuracy implementations. Often, DNNs may be deployed in tasks that do not have many existing implementations and/or implementations may be crafted by the same set of experts that are bound to have used the same methods or made the same errors.
 
\end{itemize}
\section{Preliminary Approach and Results}{\label{section:5}}
In this paper, we focus only on proposing a finer coverage criterion. An ideal coverage criterion for deep neural networks must be able to guarantee completeness, i.e., all parts of the internal decision logic of the DNN have been tested by at least one input. Recall that the value of a neuron in a particular layer in a DNN is computed as a nonlinear function of the weighted sum of neurons in the previous layer, as shown in Equation \ref{equation1}. On these lines, we propose a coverage criterion that incorporates both factors- the conditional effect of each neuron on the value of neurons in the next layer and the combinations of values of neurons in a layer \cite{KuhnCombinatorialTesting}. 

For two consecutive layers, $L_{k-1}$ and $L_{k}$ in a given (feed forward) deep neural network, let the neurons in these layers be denoted by $\{n_{1,k-1}, n_{2,k-1}, ..., n_{N_{k-1},k-1}\}$ and $\{n_{1,k}, n_{2,k}, ..., n_{N_{k},k}\}$ respectively, where $N_{k}$ denotes the total number of neurons in $L_{k}$. For any test input $t$, a neuron $n$ is said to be activated if its value is greater than a certain threshold, for example, 0. Formally, if $\phi(t,n)$ denotes the activation of neuron $n$ when the input to the deep neural network is $t$, then if $\phi({t,n})$ \textgreater 0 (or any other threshold value, depending on activation function) then the neuron is said to be activated or \textit{fired}. Therefore, for a given neuron $n$ and test input $t$, the condition $\phi(t,n)$ $\textgreater$ 0 can have two values, true or false, depending on whether the neuron is activated or not. 

Based on these definitions, our coverage criterion is defined as the 2-way coverage \cite{KuhnCombinatorialTesting} for every such triplet in the DNN: ($n_{i,k-1}$, $n_{j,k-1}$, $n_{q,k}$). Formally, for a given test set for \textit{n} variables, simple \textit{t-way} combination coverage is the proportion of \textit{t-way} combinations of \textit{n} variables for which all variable-values configurations are fully covered. By ensuring 2-way coverage on three such distinct neurons, we are able to cover, (1) the independent effect a condition (activation of neuron in $L_{k-1}$) has on an outcome (value of neuron in the next layer, $L_{k}$), (2) the failures that may arise because of the `interaction' or activation values of neurons in the same layer $L_{k-1}$. While the first coverage is more inspired by MC/DC\cite{Hayhurst2001ACondition} and other traditional software-coverage criteria, the second coverage is inspired by combinatorial testing \cite{KuhnCombinatorialTesting}. This kind of testing is based on the fact that not every parameter contributes to every failure, and empirical data suggest that \emph{nearly all failures are caused by interactions between relatively fewer parameters}. This finding has important implications for testing because it suggests that testing combinations of (fewer) parameters can provide highly effective fault detection. In our scenario, since values of multiple (but not always all) neurons in the previous layer contribute towards the values of neurons in the next layer, such a method is able to test for multiple values that a condition (weighted sum in Equation \ref{equation1}) can take, which is also one of the requirements for traditional software coverage criteria such as MC/DC\cite{Hayhurst2001ACondition}. 

For preliminary results, we approach guided test input generation via joint optimization\cite{Pei2017DeepXplore:Systems}. Any triplet not having achieved 100\% coverage is randomly chosen to determine which combination(s) of activation values has not been covered. Consider, for example, the DNN instance where n\textsubscript{i,k-1} is fired but n\textsubscript{j,k-1} is not activated. The decision neuron n\textsubscript{q,k} is fired. The objective becomes  
\begin{equation}
    F_{n,t} = f_{n_{i,k-1}}(t) + f_{n_{j,k-1}}(t) + f_{n_{q,k}}(t), 
\end{equation}
where 
\begin{itemize}
    \item  $f_{n_{i,k-1}}(t)$ = $\phi({t},{n_{i,k-1}})$ needs to be maximized such that $\phi({t},{n_{i,k-1}})$ $\textgreater$ 0 (or the decided threshold), 
    \item $f_{n_{j,k-1}}(t)$ = $\phi({t},{n_{j,k-1}})$ needs to be minimized such that $\phi({t},{n_{j,k-1}})$ = 0, and
    \item $f_{n_{q,k}}(t)$ = $\phi({t},{n_{q,k}})$ needs to be maximized such that $\phi({t},{n_{q,1}})$ $\textgreater$ 0. 
\end{itemize} 
The objective function to maximize coverage therefore becomes,
\begin{equation}
    F_{{n,t}} = \phi({t},{n_{i,k-1}}) - \phi({t},{n_{j,k-1}}) + \phi({t},{n_{q,k}}) .
\end{equation}
 Because the individual terms in $F_{{n,t}}$ are activation values of certain neurons in certain layers and $\phi(\textit{t},\textit{n})$ for any $n$ is a sequence of stacked functions, the gradient  $\pdv{F\textsubscript{\textit{n}}(\textit{t})}{\textit{t}}$ can be calculated using the chain rule in calculus, i.e., by computing layer-wise derivatives backwards from the layer containing neuron $n$ until reaching the input layer which takes input $t$ \cite{Pei2017DeepXplore:Systems}.Hence, the input $t$ can be manipulated in steps to maximize $F_{{n,t}}$. The oracle we use is the same as \cite{Pei2017DeepXplore:Systems}, so the second objective is to generate differential behavior causing inputs.   

We evaluated our coverage metric on three DNNs that classify the MNIST dataset of handwritten digits: LeNet-1, LeNet-4 and LeNet-5. Since the primary goal of our work is to introduce and test a more fine-grained coverage metric, our test input generation method and oracle share the limitations mentioned in section \ref{section:4}.
The metrics used for determining the validity of our coverage criterion were:
\begin{itemize}
    \item The coverage obtained on ten random test inputs, and
    \item The ratio of number of corner cases found to the number of total test inputs.
\end{itemize}
Ideally, the coverage for ten random test inputs (not generated using a guided method) must be low, i.e., the criterion must be difficult to achieve for random inputs, and the adversarial ratio must be high. We currently use multiple implementations \cite{Pei2017DeepXplore:Systems} as an oracle, introduced in \ref{section:4}, and only one image manipulation, brightness. The results are summarized in Table \ref{Tab:2}\footnote{Our implementation involves trying to optimize for image manipulations to generate differentially behaving, more coverage test inputs from \textit{all} inputs, whether or not they cause differential behavior when not manipulated at all.}.
\begin{table}[]
    \centering
    {\def\arraystretch{1.5}
    \begin{tabular}{|c|c|}
         \hline
         Metric & Result \\
         \hline
         Coverage for 10 random inputs & 8.9\% \\
         \hline
         Guided coverage for 550 test inputs & 31\% \\
         \hline
         Number of corner case behaviors found for 550 inputs tested & 483 \\
         \hline
         Adversarial Ratio & 87.8\% \\
         \hline
    \end{tabular}}
    \caption{Evaluation of coverage metric on LeNet architectures for MNIST dataset. All results are an average over LeNet-1, LeNet-4, LeNet-5.}
    \label{Tab:2}
    \vspace{-4mm}
\end{table}
We then compared our proposed coverage criterion with existing coverage metrics. We found that for the same dataset and DNNs, the average neuron coverage\cite{Pei2017DeepXplore:Systems} for ten random test inputs over the three DNNs is 30.5\% (threshold used is 0), as opposed to 8.9\% for our coverage criterion. Further, for LeNet-1, 100\% neuron coverage can be achieved with just two corner-case inputs and a lot of corner-case inputs can be found beyond achieving 100\% neuron coverage. On the other hand, LeNet-1 achieves close to 11.6\% coverage for our criterion over 550 test inputs. This is because the most common activation pattern in the DNN for the given test inputs is all neurons being fired/activated, and hence 2-way coverage is difficult to achieve. 
The average neuron coverage across all three DNNs using guided test input generation is 98.5\% for 550 test inputs, but is 31\% for our proposed coverage criterion using the same test input generation method. The maximum adversarial ratio obtained using DeepCover \cite{Sun2018TestingNetworks} for DNNs of similar size is 11\%. Similarly, DeepCT\cite{Ma2018CombinatorialSystems} achieves less than 10\% adversarial ratio for a DNN of similar size, for 10,000 test inputs. These results confirm that for testing DNNs, it is important to have a more fine-grained coverage metric that not only incorporates inter-layer relationships, but also the relative activations of neurons in the same layer. While the large number of triplets may seem like a computational bottleneck, the average time taken to update coverage for LeNet-5 with the most number of triplets (651720) is 2.08 seconds. 

\section{Conclusion}\label{section:6}
The absence of a transparent decision logic makes it impossible to apply traditional software testing methods to DNNs. This paper examines existing testing methods for deep neural networks and recognizes several limitations such as coarse coverage criteria, open ended processes, unreliable oracles, inefficient test input generation methods, inability to scale to larger DNNs and different network architectures, etc. Further, we propose a fine-grained coverage criterion for feed forward DNNs that takes into account the condition-decision relationships between adjacent layers and the combinations of values of neurons in the same layer. A set of ten random test inputs could only achieve 8.9\% of our coverage criterion. Further, when coupled with gradient-based search techniques and multiple implementations oracle, it is able to achieve an average 87.8\% adversarial ratio over three models. The ability to test the internal logic of a DNN to a greater extent makes its performance better than existing methods. The scalability of the coverage method to larger-sized real-world DNNs and its adaptation to different network architectures is yet to be tested.

\section{Acknowledgements}
This material is based upon work supported in part by the National Science Foundation under Grant No. CNS: 1650512, conducted in the NSF UICRC Center of Visual and Decision Dynamics. 
This research was also supported by the Northrop Grumman Mission Systems’ University Research Program.

\bibliographystyle{plain}
\bibliography{main.bib}

\end{document}